\documentstyle[prl,epsfig,aps,multicol]{revtex}
\tighten
\renewcommand{\narrowtext} 
{\begin{multicols}{2}\global\columnwidth21pc} 
\renewcommand{\widetext} 
{\end{multicols}\global\columnwidth42.5pc} 
\newcommand{\be}{\begin{eqnarray}}
\newcommand{\ee}{\end{eqnarray}}
\def\beq{\begin{equation}}
\def\eeq{\end{equation}}
\begin{document} 
\draft 
\title{Classical intermittency and quantum Anderson transition} 
\author{Antonio M. Garc\'{\i}a-Garc\'{\i}a} 
\address{Laboratoire de Physique Th\'eorique et Mod\`eles 
Statistiques, B\^at. 100, \\ Universit\'e de Paris-Sud, 
91405 Orsay Cedex, France}
\maketitle
\begin{abstract}
We investigate the properties
 of quantum systems whose classical counterpart presents intermittency.
 It is shown, by using 
 recent semiclassical techniques, that the
 quantum spectral correlations of such systems are expressed
 in terms of the eigenvalues of an anomalous diffusion operator. 
For certain values of the parameters leading to ballistic diffusion and 
 $1/f$ noise
 the spectral properties of our model  
 show similarities
 with those of a disordered system at the Anderson transition.
 In Hamiltonian systems, intermittency
 is closely 
 related to the presence of cantori in the classical phase space. 
 We suggest, based on this relation, that our findings may be relevant for the description of 
 the spectral correlations of (non-KAM) Hamiltonians  
 with a classical phase space homogeneously filled by cantori. Finally we discuss
the extension of our results to  
higher dimensions and their relation to Anderson models with long range hopping.

\end{abstract}

\pacs{PACS numbers: 72.15.Rn, 71.30.+h, 05.45.Df, 05.40.-a} 
\narrowtext

 The quantum properties of a disordered system, namely, 
a non interacting particle in a random potential are strongly
 affected by both the dimensionality of the space and 
the strength of disorder.
 In less than three dimensions the wavefunctions are localized in the 
 thermodynamic limit for any amount of disorder.   
  In three and higher dimensions there exists
 a metal insulator transition (MIT) for   
 a critical amount of disorder. Thus for disorder below the critical one
 the wavefunctions are extended through the sample,
 the Hamiltonian is accurately approximated by a random 
 matrix with the appropriate symmetry, and the 
 spectral correlations are given 
 by Wigner-Dyson (WD) statistics \cite{mehta}.  In the opposite limit, wave functions
 are exponentially localized and the spectral correlations are described 
by Poisson statistics. 
 A similar situation occurs in 'quantum chaos'. 
 The celebrated Bohigas-Giannoni-Schmit
 conjecture \cite{oriol} states that the WD statistics
 applies to the spectral correlations 
of quantum systems whose classical counterpart is chaotic. On the other 
hand, it is broadly accepted \cite{tabor} that Poisson statistics
 describes the spectral correlations of quantum systems  
whose classical counterpart is integrable. 
  
  Deviations from WD statistics due to  
 wavefunction localization has been intensively investigated
 in recent years. 
 In disordered systems, they are expressed through the dimensionless
 conductance $g=E_c/\Delta$ 
 where $E_c=\hbar/t_c$ is the Thouless energy, $t_c$ is the classical
 time to cross the sample diffusively 
 and $\Delta$ is the quantum mean level spacing. In the metallic
 regime $g \rightarrow \infty$ and WD statistics applies. 
  Non perturbative corrections due to a finite $g \gg 1$ were recently
  evaluated by Andreev and Altshuler \cite{andre}
  in the framework of the supersymmetry method \cite{efetov}. They managed to express the 
 two level spectral correlation function in terms of 
the spectral determinant of the classical diffusion operator.  
  In the context of quantum chaos deviations from WD statistics 
 are expected  
due to the non universality of 
 short periodic orbits (here we do not discuss other sources of 
deviation as dynamical localization or a mixed phase space).
 In a recent development \cite{ogam}, the two level spectral 
 correlation function encoding such deviations was found to be
  related to the spectral determinant
 of the classical Perron-Frobenius operator which controls the evolution of 
the classical phase space. Unfortunately explicit results
 are hard to obtain since there is no a general recipe to compute 
the eigenvalues of this operator for a generic classical Hamiltonian.

As disorder strength further increases $g \sim 1$
 localization effects become dominant and  eventually
 the system undergoes a MIT. 
 At the MIT, the wavefunction moments $P_q$
  present anomalous scaling with respect 
  to the sample size \cite{wegner} $L$, 
$P_q=\int d^dr |\psi({\bf r})|^{2q}\propto L^{-D_q(q-1)}$,
where $D_q$ is a set of exponents describing the transition. 
Wavefunctions with such a non trivial 
scaling are said to be multifractal 
(for a review see \cite{janssen}). 
 Spectral fluctuations at the MIT (commonly referred as 
'critical statistics \cite{kravtsov97}) 
are intermediate between 
 WD and Poisson statistics. 
 Typical features include: scale invariant spectrum \cite{sko},
 level repulsion and sub-Poisson number 
 variance \cite{chi}. 
Different generalized random matrix model have been successfully 
employed to describe 'critical statistics' \cite{Moshe}.

A natural question to ask is whether 
 critical statistics
is related to any kind of classical 
 motion. We shall show that, for a certain range 
 of parameters, the spectral correlations of quantum systems whose classical 
 counterpart presents intermittency \cite{pome} and $1/f$ noise  
 are described by critical statistics. This is the main result
 of this work. We also suggest 
that our results may be useful to describe 
the spectral correlations of non-KAM  Hamiltonians with a classical 
phase space homogeneously filled by cantori.  
The organization of the paper is as follows: 
Classical intermittency is introduced  
by studying the evolution of a simple non linear map. In the context of Hamiltonian systems,  we also 
 discuss its relation with classical phase space structures
  Quantum spectral correlations associated with classical intermittency
 are then investigated by using 
 the semiclassical Andreev-Altshuler formalism above mentioned. 

 Finally we discuss the extension of our results to higher dimensions 
and its relation with Anderson models with long range hopping.  
\section{Classical intermittency}
 The phenomenon of intermittency is characterized by 
 long periods of laminar 
(regular) motion interrupted by short irregular bursts \cite{pome}. 
As a simple example of a dynamical system with such features we 
 investigate the following map on the real line \cite{pome,geisel}:
$x_{n+1}=f(x_n)$, where $f(x)$ verifies $f(x)=-f(-x)$ and 
$f(x+N)=f(x)+N$ with $N$ an integer. With the above rules the 
 map needs to be defined only in a restricted interval,   
$f(x_n) =  (1+\epsilon)x_n + a|x_n|^z-1, 0<x_n<1/2$, with $z$ and $a$ 
real numbers and $\epsilon \rightarrow 0$ is a small 
 control parameter utilized to set the scale of the laminar phase. 
  The laminar motion has its origin
 at  points $x_n \sim N$ where $x_{n+1}\sim x_n \pm 1$ and thus the orbit
 is transfered to the 
 same position in the neighboring cell. In a continuous time 
this corresponds to  ballistic motion. Eventually  
 the orbit leaves the region $x \sim N$ and the dynamics becomes
 chaotic. The duration of the chaotic phase 
 is typically much shorter than the ballistic one. 
We mention that
 due to universality \cite{schuster}, intermittency appears for any $f(x)$ with 
 a Taylor expansion for $x \ll 1$ given by the above relation.
 In \cite{geisel} it was found that   
 the density of probability of staying in the laminar phase a time $t$
 (or a distance $r$) has a power law tail,
\be
\label{lev}
\psi(t,r)\sim \frac{{\hat b}}{(1+r)^{\mu}}\delta(|r|-{\hat b}t)
\ee
where ${\hat b} = 2^z a+\epsilon/2$ and  $\mu=\frac{z}{z-1}$. Indeed
 the motion is  
superdiffusive \cite{geisel} for $2< \mu < 3$ and ballistic for $1 < 
 \mu \le 2$. 
Based on the above result, 
 Zumofen and Klafter \cite{zum} calculated the probability $P(r,t)$ to be
 at location $r$ at time $t$ 
in the framework of the continuous random walk model,
\be
\label{fp}
P(s,k)\sim \frac{1}{is+{b}|k|^{\mu-1}}~~~2 \le \mu \le 3 \\     
\frac{s^{\mu-2}}{is^{\mu-1}+{b}|k|^{\mu-1}} \nonumber   
 ~~~1 < \mu < 2
\ee
where $P(s,k)$ is the Fourier transform of $P(x,t)$ and $b \sim {\hat b}$ 
up to factors of order the unity. 
For $2 < \mu \le 3$, the above equation is a solution of the following fractional 
 Fokker Planck equation (for a recent review see \cite{rev}), 
\be
\frac{\partial P(x,t)}
{\partial t}- \frac{b}{2}\frac{\partial^{\mu-1}}{\partial 
|x|^{\mu -1}}P(x,t)= \delta(x)\delta(t).
\ee  
where we define the fractional derivative as in \cite{rev}.

The phenomenon of intermittency in Hamiltonian systems 
has been related \cite{meiss,geisel1} to the 
 presence of cantori \cite{cantori} in phase space. 
 A trajectory is typically trapped 
 in the intricate cantori structures 
for times 'long' as compared with those 
involved in the transport through the pure chaotic phase space.  
Meiss and coworkers \cite{meiss} described the  
 transport through cantori in phase space 
 as a random walk in a Bethe lattice. 
 Such simplified model predicts a power law tail 
for the waiting times inside a cantori in agreement with 
numerical simulations \cite{klaf}.  
 Details of the distribution as the decay exponent
 are sensitive to the scaling properties 
 of the cantori which depend on the considered energy 
and are thus non universal. Since    
 waiting times in phase space lead to ballistic motion in real space 
\cite{geisel1}, a trajectory in real 
 space is a mixture of long ballistic flights governed 
 by a power-law distribution eventually interrupted by random 
 walks when the trajectory escapes from the cantori region.    
The above simplified model cannot in principle describe the full
 complexity of typical KAM Hamiltonians with a mixed phase space 
 where cantori with different scaling properties coexist 
with pure chaotic and integrable components 
(for a recent investigation in this direction see \cite{klaf1}).
A different situation occurs in Hamiltonians in which, due to the
 non analyticity of the classical potential, the KAM theorem does not 
 hold. In certain cases \cite{levitov97}, as a parameter is switched on,
the whole classical phase space undergoes an abrupt transition from integrable 
 to homogeneously filled with cantori. 
The absence of 
additional 
classical structures permits in this case utilize the 
 formalism above explained though   
the details of $P(r,t)$ may 
depend on the considered energy. 
Others class of systems with similar properties is that 
of pseudo-integrable billiards \cite{shudo}, the 
phase space is also fractal and the 
classical motion presents anomalous transport properties \cite{artuso,bogo}.

\section{Quantum manifestations of classical intermittency.}
 We now investigate quantum
 manifestations of classical intermittency by using semiclassical techniques.  
 Our starting point is the study of the connected two level correlation function,
\be
R_{2}(s,g)=\Delta^2 \langle \rho(\epsilon -s/2)\rho(\epsilon +s/2) \rangle -1 
\ee
where $\rho(\epsilon)$ is the density of states at energy $\epsilon$, $\Delta$ is the 
 mean level spacing, 
the energy $s$ is expressed in units of $\Delta$ and 
the averaging is over an ensemble for disordered systems
  and over an interval 
 of energy for single deterministic chaotic systems. The
  spectral properties depend on the ratio $g=E_{c}/\Delta$.
In the ergodic limit $g \rightarrow \infty$, 
the Hamiltonian can be accurately 
 approximated by a random matrix with the appropriate symmetry 
  and WD statistics applies. 
For instance,    
 for broken time reversal invariance, 
$R_2(s,0)= - \frac{\sin^2(\pi s)}{\pi^2s^2}$.  
 Perturbative corrections ($g \gg 1$ and $s \gg 1$) to this 
 result are evaluated \cite{chi} by simple perturbation
 theory, 
\be
R_{2}^{pert}(s,g)=\frac{1}{2\pi^2}Re\sum_n P^2(\epsilon_n,s) \nonumber 
\ee
where $P(\epsilon_n,s)$ is the 
 propagator of the diffusion equation (in our case (\ref{fp})), $\epsilon_n$
 are the eigenvalues of the diffusion equation and $n$ runs over the 
 integers. As expected, the lowest mode $n = 0$ reproduces the asymptotic form of the WD statistics.
 Non perturbative corrections (leading to the oscillating terms) to this result
have been worked out by using the supersymmetry method 
 \cite{andre}. This approach is valid only in the $s \gg 1$ limit except 
 for systems with broken 
time reversal invariance where it is supposed to hold for
 any $s$. In this case,  
\be
\label{kt}
R_2(s,g) = R_{2}^{pert}(s,g)+R_{2}^{osc}(s,g)\\ 
R_{2}^{osc}(s,g)= \cos(2\pi s)\,\frac{D(s,g)}{2s^2\pi^2} \nonumber
\ee
 where $D(s,g)$ is the 
 spectral determinant associated to the diffusion equation.  
 For normal diffusion 
$\small{D^{-1}(s,g)= \prod_{n \neq 0}(1+\frac{s^2}{\epsilon_{n}^2})}$
 and $\epsilon_{n}=gn^{2}$. 
In principle one may 
ask whether the above formalism is applicable 
to the case of anomalous diffusion. 
It turns out that for $\mu \ge 2$ this can be demonstrated by mapping it 
onto an Anderson model with long range disorder \cite{prbm}(see below). 
For $\mu < 2$, although the mapping is in principle possible, we are not aware of a rigorous proof so our results should be considered in this case as a conjecture.        
 
The spectral correlations associated with classical 
intermittency are now studied by using the above semiclassical techniques. 
We evaluate (\ref{kt}) for different $\mu$ and then
 we investigate the long range spectral correlations  
 by the analysis of the number variance. 
We recall that 
 the number variance $\Sigma^{2}(L)= \langle L \rangle^2 - \langle L^2 \rangle=L+
2\int_{0}^{L}ds(L-s)R_2(s,g)$
  measures
the stiffness of the spectrum. In the 
 metallic regime, for eigenvalues 
 separated less than the Thouless energy, fluctuations are small 
  and $\Sigma^{2}(L) \sim \log(L)$ for $L \gg 1$. Beyond 
 the Thouless energy spectral fluctuations get stronger and
 $\Sigma^{2}(L)\sim L^{d/2}$ where $d$ is the dimensionality of the space.
For disorder strong enough eigenvalues are
uncorrelated (Poisson statistics) and  $\Sigma^{2}(L)=L$. At the MIT, the number variance is asymptotically proportional
to $\chi L$ ($\chi < 1$) with $\chi$ being related with the multifractal
 scaling of the wavefunction moments \cite{krav2}. 
\subsection{Case I: $1 < \mu < 2$}
 This case corresponds with classical ballistic diffusion $\langle r^2 \rangle \propto t^2$ 
\cite{geisel}. 
The propagator of the diffusion equation is given by (\ref{fp}),
 $P(\epsilon_n,s)=\frac{s^{\mu-2}}{is^{\mu-1}+\epsilon_n}$, 
$D^{-1}(s,g)= \prod_{n \neq 0}(1+\frac{s^{2\mu-2}}{\epsilon_{n}^2})$,
$\epsilon_{n}= g|n|^{\mu -1}$ (where periodic boundary condition and assumed) 
and, by using (\ref{kt}), 
$R_{2}(s,g) \sim g^{(\mu-1)^{-1}}/s$ for $s \gg 1$. 
 The parameter $b$ in (\ref{fp}) is related to $g$ as follows.
For normal diffusion $E_c \sim b/L^2$ and $g \sim bL^{d-2}$. However,  
  in our case, since the diffusion is ballistic,
 $E_c \sim b^{2\mu-3}/L$ and  $g \sim b^{2\mu -3}$ for $1 < \mu <  2$.
  We remark that the scale invariance of $g$ may be modified 
by quantum corrections as in the case of a two dimensional
 weekly disordered conductor. The number variance behaves asymptotically as 
  $\Sigma^{2}(L) \sim g^{(\mu-1)^{-1}}L\log L$ with a subleading 
 linear term as at the MIT. Additional work on the 
 wavefunctions is needed to fully
 characterize the quantum properties in this region.   
\subsection{ Case II: $\mu = 2$}
In this case,
$D^{-1}(s,g)= \prod_{n \neq 0}(1+\frac{s^{2}}{n^2g^2})=\frac{1}{g^2}
\frac{\pi^2 s^2}
{\sinh^2(\pi s/g)}$ and 
 $R_2(s,g)$ can be explicitly evaluated,
\be
R_2(s,h)= h^2\frac{\sin^2(\pi s)}{\sinh^2(\pi hs)} 
\ee            
 where $h=1/g \ll 1$. Remarkably, this correlation function has been
 put forward as a definition of critical statistics \cite{kravtsov97}. 
It reproduces 
 typical features at the MIT as  
 level repulsion (for $s \ll 1$ coincides with Wigner-Dyson statistics)
 and sub-Poisson number variance 
($\Sigma^2(L) \sim hL$ for $L \gg 1$). In this case
  $h \sim 1/b$ is also  
scale invariant. As discussed below, 
by mapping this case onto a Anderson model 
with long range disorder one can show that the wavefunctions are indeed multifractal
 as at the MIT. Finally, we remark that the classical motion associated to $\mu=2$
 leads to $1/f$ noise \cite{geisel1}.

\subsection{ Case III: $2 <\mu \le 3$}

Now the dynamics is superdiffusive but sub-ballistics, $\langle r^2 \rangle 
\sim t^{4-\mu}$ \cite{klaf}. Classical anomalous 
diffusion is described by the propagator, 
$P(\epsilon_n,s)=\frac{1}{is+\epsilon_n}$, 
$D^{-1}(s,g)= \prod_{n \neq 0}(1+\frac{s^{2}}{\epsilon_{n}^2})$ with 
$\epsilon_{n}= g|n|^{\mu-1} $. 
 The asymptotic behavior of $R_2(s,g) 
\sim s^{-2+1/(\mu-1)}$ is power low instead of exponential. The conductance 
 $g$ is scale dependent and decreases as the system size increases,  
 $E_c \sim L^{-2/(4-
\mu)}$, $g \sim L^{1-2/(4-\mu )}$.
For $s \gg g$, 
the power law tail of $R_2(s,g)$ 
 leads to
  $\Sigma^2(L) \sim  L^{1/(\mu-1)}$. 
 Both the scaling 
 of $g$ and the spectral properties   
 resemble those of a   
 weakly disordered conductor in $d= \frac{2}{{\mu -1}}$ \cite{prbm} 
 dimensions.  
 Finally for $\mu=3$ one recovers the expected behavior
 $\Sigma^2(L) \sim \chi \sqrt{L}$ of a 1D weakly 
disordered conductor (no  anomalous  diffusion). 

 We mention 
 that similar findings have been reported
 in Anderson models with long range $1/r^{\mu-1}$ 
 hopping \cite{prbm}. 
  For $2 \le \mu \le 3$  
the classical transport is also described by 
 (\ref{fp}) provided that the disorder is weak enough \cite{fog}.
By using the supersymmetry method not only the spectral correlations but 
also the wavefunctions can be studied analytically. 
 It turns out that the eigenfunctions are power law localized 
 \cite{prbm} $|\psi(r)| \sim r^{-\mu +1}$. 
In the thermodynamic limit, 
they become localized for $\mu > 2$, delocalized for $\mu < 2$ and multifractal for $\mu = 2$ 
 as at the MIT. 
Finally, we point out that typical features of these long 
 range hopping models as 
 power law localization and
 criticality also appears in higher dimensions \cite{prbm}. 
 Thus, in $d$ dimensions, wavefunctions
 are power-law localized for any exponent $\mu$ and  
 multifractal for $\mu = d + 1$.

We now discuss 
under what conditions the above findings are relevant for deterministic Hamiltonians. Obviously a first constraint is that the classical dynamics 
be described by (\ref{fp}). As discussed previously this could be the case
for non KAM systems \cite{levitov97} with a classical 
phase space homogeneously filled by cantori.             
Although the parameters defining (\ref{fp})  
 may depend on the considered energy,   
 numerical results \cite{geisel1} suggest that 
 the in certain cases they are 
  barely modified in a broad window of energy and are close to the ones leading to 
  $1/f$ noise.    
 From a quantum mechanical point view 
 our results are only valid in the limit 
 $g \gg 1$ where interference (not tunneling) is the dominant 
 quantum feature. 
In the context of quantum chaos this scale
  corresponds with ${\hat g} \sim \Delta W/\hbar$ 
where $\hbar$ is the Planck constant and $\Delta W$ is 
the flux swept across the cantorus with less flux at a given energy 
in one iteration of the 
 map. Our results are thus applicable for energies such that ${\hat g} \gg 1$. 
 This limit corresponds to the case when quantum mechanics
 can resolve the classical barrier. For ${\hat g} \sim 1$ quantum mechanics 
 cannot resolve the gap and in order to cross it must tunnel through it.    
 We mention that recently it has been reported \cite{ant3} 
 that the high energy excitations  
of non-KAM systems as the Anisotropic Kepler problem or the Coulomb  billiard 
 \cite{levitov97} are correlated 
 according to critical statistics. It would be interesting to check whether the classical mechanics
 of these systems present $1/f$ noise as predicted in this letter.
          
In conclusion, we have investigated quantum manifestation of classical intermittency. 
It has been shown that for classical
 ballistic 
 diffusion and $1/f$ noise (a special case of intermittency)
 the spectral correlations are given by critical statistics as at the MIT. In other cases
 the classical motion 
is superdiffusive but sub-ballistic, the wavefunctions are 
power law localized and the spectral correlations
 are similar to those of a weakly disordered conductor in less 
than two dimensions.   
In the context of Hamiltonian systems 
we have suggested that these results may be relevant for the description of the spectral correlations of non KAM systems with classical phase space homogeneously filled by cantori.          

Discussions with Patricio Leboeuf, Yan Fyodorov and Vladimir Kravtsov are 
gratefully acknowledged. This work was supported by the European Union
  network ``Mathematical aspects of quantum chaos''.

\end{multicols}
\end{document}